\documentclass[twocolumn, superscriptaddress,amsmath, prl, aps]{revtex4-2}
\usepackage[utf8]{inputenc}
\usepackage{times,color,amsthm,graphics,graphicx,bm,bbm,dcolumn}
\usepackage{epsfig}
\usepackage{graphicx}
\usepackage{siunitx}
\usepackage{xcolor}
\usepackage[colorlinks,urlcolor=blue,citecolor=blue,linkcolor=teal]{hyperref}

\usepackage{lineno}
\definecolor{red}{RGB}{0,0,0}

\usepackage{amsmath}
\def \mum{\mu \mathrm{m}}

\def \mHz{\mathrm{Hz}}
\def \mnK{\mathrm{nK}}
\def \mms{\mathrm{ms}}

\setcitestyle{super}

\begin{document}

\title{Shapiro steps in strongly-interacting Fermi gases}



\author{G.~Del Pace}
\email[E-mail: ] {delpace@lens.unifi.it}
\affiliation{Department of Physics, University of Florence, 50019 Sesto Fiorentino, Italy}
\affiliation{European Laboratory for Nonlinear Spectroscopy (LENS), University of Florence, 50019 Sesto Fiorentino, Italy}
\affiliation{Istituto Nazionale di Ottica del Consiglio Nazionale delle Ricerche (CNR-INO) c/o LENS, 50019 Sesto Fiorentino, Italy}

\author{D.~Hern\'andez-Rajkov}
\affiliation{European Laboratory for Nonlinear Spectroscopy (LENS), University of Florence, 50019 Sesto Fiorentino, Italy}
\affiliation{Istituto Nazionale di Ottica del Consiglio Nazionale delle Ricerche (CNR-INO) c/o LENS, 50019 Sesto Fiorentino, Italy}

\author{V.~P.~Singh}
\affiliation{Quantum Research Centre, Technology Innovation Institute, Abu Dhabi, UAE}

\author{N.~Grani}
\affiliation{Department of Physics, University of Florence, 50019 Sesto Fiorentino, Italy}
\affiliation{European Laboratory for Nonlinear Spectroscopy (LENS), University of Florence, 50019 Sesto Fiorentino, Italy}
\affiliation{Istituto Nazionale di Ottica del Consiglio Nazionale delle Ricerche (CNR-INO) c/o LENS, 50019 Sesto Fiorentino, Italy}

\author{M.~Fr\'ometa Fern\'andez}
\affiliation{European Laboratory for Nonlinear Spectroscopy (LENS), University of Florence, 50019 Sesto Fiorentino, Italy}
\affiliation{Istituto Nazionale di Ottica del Consiglio Nazionale delle Ricerche (CNR-INO) c/o LENS, 50019 Sesto Fiorentino, Italy}

\author{G.~Nesti}
\affiliation{European Laboratory for Nonlinear Spectroscopy (LENS), University of Florence, 50019 Sesto Fiorentino, Italy}
\affiliation{Istituto Nazionale di Ottica del Consiglio Nazionale delle Ricerche (CNR-INO) c/o LENS, 50019 Sesto Fiorentino, Italy}

\author{J.~A.~Seman}
\affiliation{Instituto de Física, Universidad Nacional Autonoma de Mexico, C.P. 04510 Ciudad de Mexico, Mexico}

\author{M.~Inguscio}
\affiliation{Department of Engineering, Campus Bio-Medico University of Rome, Rome, Italy}
\affiliation{European Laboratory for Nonlinear Spectroscopy (LENS), University of Florence, 50019 Sesto Fiorentino, Italy}
\affiliation{Istituto Nazionale di Ottica del Consiglio Nazionale delle Ricerche (CNR-INO) c/o LENS, 50019 Sesto Fiorentino, Italy}

\author{L.~Amico}
\affiliation{Quantum Research Centre, Technology Innovation Institute, Abu Dhabi, UAE}
\affiliation{INFN-Sezione di Catania, Via S. Sofia 64, 95127 Catania, Italy}
\affiliation{Dipartimento di Fisica e Astronomia, Universit\`a di Catania, Via S. Sofia 64, 95123 Catania, Italy}

\author{G.~Roati}
\affiliation{European Laboratory for Nonlinear Spectroscopy (LENS), University of Florence, 50019 Sesto Fiorentino, Italy}
\affiliation{Istituto Nazionale di Ottica del Consiglio Nazionale delle Ricerche (CNR-INO) c/o LENS, 50019 Sesto Fiorentino, Italy}

\begin{abstract}
We report the observation of Shapiro steps in a periodically driven Josephson junction between strongly-interacting Fermi superfluids of ultracold atoms. We observe quantized plateaus in the current-potential characteristics, the height and width of which mirror the external drive frequency and the junction nonlinear response.
Direct measurements of the current-phase relationship showcase how Shapiro steps arise from the synchronization between the relative phase of the two reservoirs and the external drive.
Such mechanism is further supported by the detection of periodic phase-slippage processes, 
in the form of vortex-antivortex pairs.
Our results are corroborated by a circuital model and numerical simulations, overall providing a  clear  understanding of Shapiro dynamics in atomic Fermi superfluids. Our work demonstrates phase-coherent and synchronization effects in driven strongly-interacting superfluids, opening
prospects for studying emergent non-equilibrium dynamics in quantum many-body systems  under external drives.
\end{abstract}

\maketitle
Driven many-body systems exhibit rich and complex behaviors that far extend beyond their equilibrium properties 
\cite{Pikovsky2001}.
A paradigmatic example is provided by Josephson junctions (JJs) under driving currents. A Josephson junction comprises two superconducting islands separated by a thin insulating barrier. Josephson showed that a dissipationless current $I$ can flow across the junction
, maintained solely by the phase difference $\phi$ between the two superconducting reservoirs \cite{Josephson1962}. 
Above a critical current $I_c$, the junction transitions into the so-called \textit{voltage-state}: a finite voltage $V$ develops across the junction, which is directly related to the 
phase difference dynamics as provided by 
the Josephson-Anderson relation $V=\frac{\hbar}{2e}\dot{\phi}$, $\hbar$ being the reduced Planck constant and and $e$ the electron charge. 
In this regime, the  Josephson current time evolution is characterized by the frequency $\omega_J=2e V/\hbar$. The junction is thus 
an element that operates in cycles with a natural frequency controllably set by the extrernally injected  current \cite{Barone1982, Tinkham2004}.

When the JJ is driven by a combined DC $\&$ AC bias with frequency $\omega$, 
the junction dynamics results to  be synchronized with the external drive:  $\omega_J=n \omega$,  $n$ being a positive integer. This synchronization results in quantized voltage plateaus, known as Shapiro steps, appearing at $\langle V\rangle_t = n \hbar\omega/2e$ in the time-averaged voltage-current characteristics \cite{Shapiro1963}. Shapiro steps arise from a phase-locking process where 
$\phi$ advances by $2\pi n$ during each cycle of the driving field, establishing coherence between the junction phase and the driven field.
%
Since their observation in superconducting Josephson junctions (SJJ) \cite{Shapiro1963}, Shapiro steps have been investigated in various nonlinear systems, ranging from charge density waves \cite{Grner1988}, colloidal systems \cite{Juniper2015}, superconducting nanowires \cite{Dinsmore2008}, and $^3$He weak-links \cite{Simmonds2001}. Shapiro steps in SJJs play a crucial role in metrology, calibration within electronics, and in defining quantum standards \cite{Jeanneret2009, Kohlmann2011}.

Recent years have witnessed significant progress in
developing and operating circuit architectures employing ultracold atoms \cite{Amico2022}. 
These 
 \textit{atomtronics} circuits feature enhanced control and flexibility of their working conditions, 
that may go beyond conventional electronics. 
One of the most promising atomic platforms for implementing such circuits consists of ultracold Fermi gases \cite{polo2024perspective}. 
These systems provide the unique opportunity to realize different regimes of superfluidity, ranging from the Bardeen-Cooper-Schrieffer (BCS) limit of weakly-bound fermion pairs to a Bose-Einstein condensate (BEC) of tightly bound molecules, including the intermediate universal, strongly-correlated unitary regime \cite{MakingProbingFermi,2012}. 
In particular, unitary Fermi gases (UFG) serve as prototypical examples of strongly-interacting systems, sharing similarities with other strongly-interacting Fermi systems found in nature, spanning a broad range of energy and length scales, from nuclear matter to neutron stars \cite{2014}.
Research on quantum transport phenomena within these systems has explored a variety of configurations, including mesoscopic two-terminal setups \cite{Krinner2017-iy}, ring geometries \cite{Cai2022, DelPace2022}, optical lattices \cite{Brown2019, Xu2019, Nichols2019, Lebrat2018}, and scenarios involving disorder \cite{Schreiber2015} or periodic modulation of the harmonic trapping potential \cite{Anderson2019,Patel2020,HernndezRajkov2021,Yan2024,cabrera2024}.

Josephson effects have been observed in both three-dimensional \cite{Valtolina2015, Kwon2020} and two-dimensional \cite{Luick2020} unitary superfluids, providing insights into the microscopic properties of these gases. However, despite various theoretical proposals \cite{Smerzi1997, Kohler2003, Eckardt2005, Grond2011}, the observation of Shapiro steps in atomic Josephson junctions has remained elusive. 
Recently, Shapiro steps have been predicted to occur by periodically oscillating the position of a thin tunneling barrier within a weakly interacting Bose-Einstein condensate \cite{Singh2023}. While 
such protocol defines new avenues for exploring Shapiro steps in atomic junctions, in particular for Fermi superfluids, it remains an open question whether strong interactions could affect their dynamics. Interactions are expected to enhance phase-coherence across the junction and potentially reduce decoherence effects associated with the finite compressibility of atomic superfluids \cite{Valtolina2015, Kwon2020}, but their role on synchronization processes is yet to be understood. Investigating these effects is crucial for deepening our understanding of strongly-interacting systems and for revealing their microscopic behavior under external drives.



\begin{figure}[t!]
\centering
\includegraphics[width=8cm]{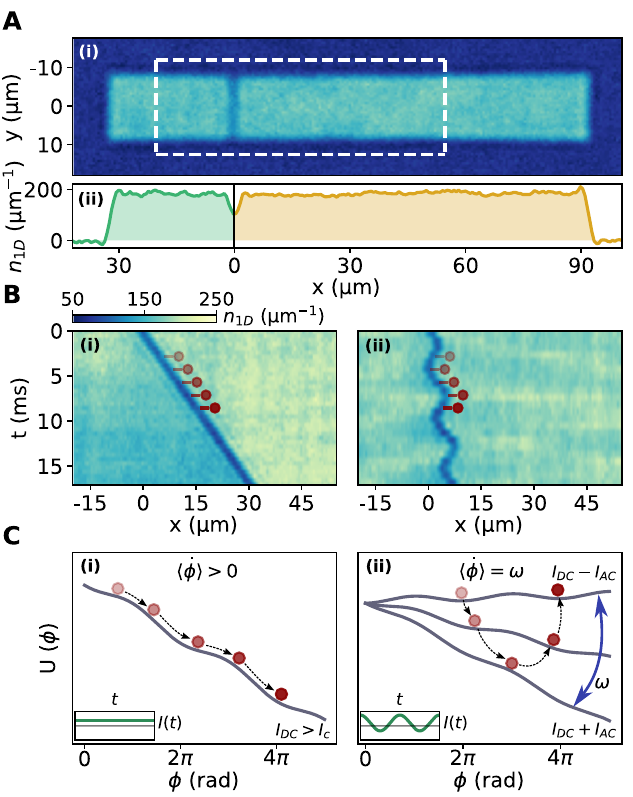}
\caption{
\textbf{Current injection in a homogeneous atomic JJ.}
(\textbf{A}) 
(\textbf{i}) \textit{In situ} image of the JJ at unitarity, averaged over $15$ repetitions. 
(\textbf{ii}) 1D density profile of the junction integrated along the y direction.
(\textbf{B}) 1D density profiles as a function of time for 
(\textbf{i}) $v_{DC} = \SI{1.5}{mm/s}$ and $x_{AC} = 0$, 
(\textbf{ii}) $v_{DC} = \SI{0.4}{mm/s}$ and $x_{AC} = \SI{2}{\mu m}$, modulating with a frequency of $\SI{175}{Hz}$.
Each row corresponds to the integrated average density profile over 5 experimental realizations. 
(\textbf{C}) Sketch of the phase particle dynamics in the washboard potential for the conditions of (\textbf{B}), with the particle transparency indicating the corresponding time.
(\textbf{i}) The voltage state of the JJ corresponds to the particle rolling down the potential, causing a voltage increase.
(\textbf{ii}) Introducing an alternating current modulates the tilt of the potential over time.
For overdamped JJ, the phase particle velocity locks to the external drive, $\langle \dot \phi \rangle_t = n\omega$, yielding to a synchronized motion that crosses $n$ potential minima during each modulation cycle. 
The insets show the current for the two illustrated cases.
}
\label{fig1}
\end{figure}

In this work, we report the observation of Shapiro steps in a periodically driven atomic Josephson junction connecting two strongly-interacting Fermi superfluids. Similar to SJJs, we find that the external driving frequency determines the height of Shapiro steps, while the driving 
amplitude affects their width. Moreover,
we access 
a detailed microscopic understanding of the physical mechanism behind the Shapiro steps by directly measuring the phase dynamics across the junction. By analyzing the current-phase relation under the external drive and observing the phase evolution across different Shapiro steps, we demonstrate 
the synchronization between the junction phase and the external drive. We observe that the Shapiro dynamics is accompanied by 
vortex-antivortex pairs emissions, in agreement with the results of Ref. \cite{Singh2023}.
Our results showcase the response of a many-body strongly-interacting system to an external drive, resulting in synchronized dynamics.



\section*{Creating and driving the atomic junction}
We produce strongly-interacting Fermi superfluids of $N \sim \SI{2e4}{}$ atom pairs  by cooling a balanced mixture of the first and third lowest hyperfine states of $^6$Li to temperatures $\sim 0.3(1)\, T_c $, where $T_c$ is the superfluid critical temperature \cite{SM}. Interparticle interactions are parametrized by $1/k_Fa$, where $k_F = \sqrt{2mE_F}/\hbar$ is the Fermi wave vector, $E_F$ the Fermi energy, and $a$ the $s$-wave scattering length. We tune $a$ in the vicinity of the Feshbach resonance at $\SI{690}{G}$ \cite{Zrn2013}, enabling the exploration of various superfluid regimes across the BEC-BCS crossover. 
In this work, we focus on two different interaction regimes: a UFG 
with Fermi energy $E_F/h = 11.8(4)\,$kHz, and a 
molecular BEC at $1/k_Fa = 3.3(1)$ with $E_F/h = 9.2(3)\,$kHz. 
The gas is confined in the $x-y$ plane using a box-like potential with dimensions $L_x \times L_y = 125 \times 17.5 \, \mu $m$^2$, and along the vertical $z$ direction 
by a strong harmonic confinement \cite{SM}. 
The trap configuration establish a nearly homogeneous in-plane atomic density with the superfluid remaining in the three-dimensional regime. To engineer the atomic JJ, we superimpose a thin repulsive optical barrier onto the atomic sample, which can be displaced along the $x$-axis, see Fig.~\ref{fig1} (A). 
The barrier intensity profile and position are dynamically controlled by a digital micromirror device (DMD), whose designed pattern is projected onto the atomic plane using a high-resolution microscope objective \cite{Kwon2020, DelPace2021}. The barrier features a FWHM
of $ \SI{0.9(1)}{\mu m}$  along the x-direction \cite{SM}, 
comparable to the healing length of the investigated superfluids, ranging between $0.6-1.7 \mu$m for typical gas parameters. The barrier height, $V_0$, is set to be larger than the gas chemical potential, $\mu_0$, with $V_0 / \mu_0= \SI{1.3(1)}{}$ for the unitary, and $\SI{1.2(1)}{}$ for the BEC JJs, ensuring that the junction operates always in the weak coupling or tunneling regime \cite{Kwon2020, DelPace2021}.

We inject currents into our junction by precisely moving the tunneling barrier along specific trajectories controlled with
the DMD \cite{Kwon2020, SM, DelPace2021}. 
The in-plane homogeneous density of the superfluid ensures that the injected current $I(t)$ is independent of its longitudinal position, and it is determined only by the instantaneous barrier velocity $v(t)$.
The instantaneous current reads $I(t) = v(t) \tilde n L_y = v(t) N/L_x$, with $\tilde n=\int n_{3D}(z) dz = N/L_x L_y$ the planar density. To introduce a biased alternating current, we move the barrier according to the trajectory $x(t) = v_{DC} t + x_{AC} \sin (\omega t)$  \cite{Singh2023}, giving rise to the instantaneous injected current 
$I(t) = I_{DC} + I_{AC} \cos(\omega t)$, with $I_{DC} = v_{DC} N/L_x$, and $I_{AC} = \omega x_{AC} N/L_x $, (Fig~\ref{fig1}B). We explore the junction operation under different driving conditions by tuning the trajectory parameters $v_{DC}$, $x_{AC}$, and $\omega$, as illustrated in Fig.~\ref{fig1} B for bias (i) and composite current injections (ii).
The junction dynamics can be portrayed with the well-known washboard potential analogy \cite{Tinkham2004}, as sketched in Fig.~\ref{fig1} C. The phase of the JJ follows the same equation of motion as that of a classical particle subject to a viscous force moving in a tilted washboard potential, where the tilt is given by the external current $I$. For a DC bias above $I_c$, the particle enters the running-phase regime, where $\langle \dot \phi \rangle_t \neq 0$, and the junction operates resistively [Fig.~\ref{fig1} C-(i)]. For an AC drive, in the overdamped regime, where friction overtakes inertia, the velocity locks to the driving tilt, resulting in a steady but synchronized descent, crossing $n$ potential wells during each modulation cycle [Fig.~\ref{fig1} C-(ii)]. This resonant motion locks the mean particle velocity to fixed values $\langle \dot \phi \rangle_t = n\omega$, leading to distinct, equally-spaced velocity steps, the Shapiro steps.


\section*{Shapiro steps in a unitary junction}

\begin{figure}[t!]
\centering
\includegraphics[width=9.1cm]{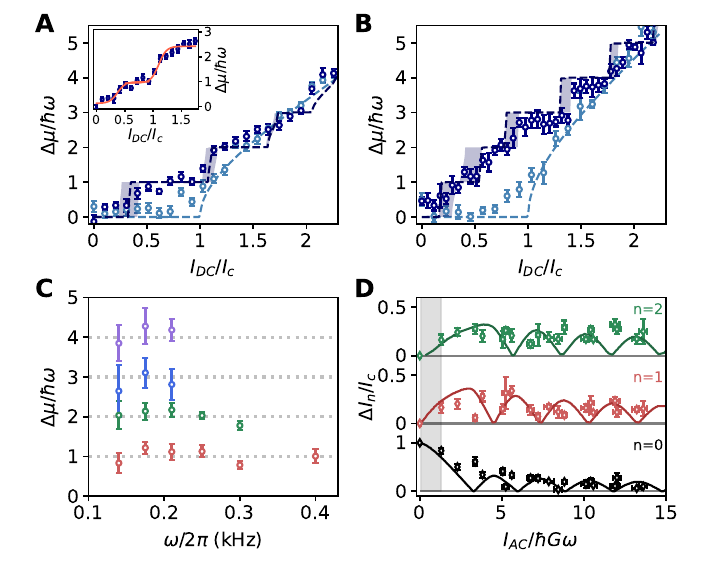}
\caption{\textbf{Shapiro steps in strongly-interacting Fermi superfluids.}
(\textbf{A}- \textbf{B}) Current-voltage characteristic for $I_{AC}/I_c=0$ (light blue), $I_{AC}/I_c= 1.6(1)$ and $\omega=2\pi \times 210 \, $Hz (dark blue)  (\textbf{A}), and for $I_{AC}/I_c= 2.4(1)$ and $\omega=2\pi \times 175 \, $Hz (\textbf{B}).
Dashed light blue lines represent the fit with the stationary solution of the undriven overdamped RCSJ model;  
dark blue ones show the results for the driven scenario, where shades account for the fitting error on $I_c$ \cite{SM}. 
Inset: phenomenological fit with $2$ independent sigmoid functions \cite{SM}.
(\textbf{C}) Shapiro step height characterization. Colors label the 1st (red), 2nd (green), 3rd (blue), and 4th (purple) step. Dotted lines represent $\Delta \mu = n \hbar \omega$. Data points represent an average over at least $3$ measurements at the same $\omega$ and different $I_{AC}$.
(\textbf{D}) Step half-width characterization: symbols report the measured step width for $175 \,$Hz, and variable $I_{AC}$. Error bars represent the error of the centroid of the multiple sigmoid fit. Solid lines correspond to the numerical solutions of the current-driven overdamped RCSJ model. $G$ results from the fit of the DC curve \cite{SM}. The gray shaded area marks the inaccessible parameter region. 
\label{fig2}}
\end{figure}

We probe the response of our junction by measuring the $I - \Delta \mu$ characteristic curves as a function of the bias current, $I_{DC}$, after three driving periods $T$ \cite{SM}, where $T= 2\pi / \omega$. Here, $\Delta \mu = \mu_L - \mu_R$ represents the chemical potential difference between the two reservoirs measured from \textit{in-situ} density images. 
Curves for $I_{AC} = 0$ and $I_{AC}>0$ for typical driving parameters are shown in Fig.~\ref{fig2} A-B. 
%
%
In the undriven scenario
, we recover the hallmark behavior of a current-biased JJ, featuring a zero-imbalance plateau for DC currents below $I_c$ \cite{Kwon2020}. 
For the AC-driven JJ instead
, we observe the emergence of distinct plateaus, with number, widths, and heights strongly depending on the driving parameters. 
To characterize the plateau properties, we perform a phenomenological fit of each step in the $I - \Delta \mu$ curve using multiple, patched sigmoidal functions \cite{SM}, from which we get the height $\Delta \mu_n$, and the width $\Delta I_n$ of each step (inset of Fig.~\ref{fig2} A).
The step height is in agreement with $\Delta \mu_n = n \, \hbar \omega$ for various driving frequencies 
independently on $I_{AC}$ (Fig.~\ref{fig2} C), as it 
is determined exclusively by the external driving frequency, while the driving strength $I_{AC}$ affects only the position and width of each step (Fig.~\ref{fig2} D).


To quantitatively describe the observed $I - \Delta \mu$ characteristics, we exploit the resistively and capacitively shunted junction (RCSJ) circuital model \cite{Barone1982, Tinkham2004}. In this model, the junction is set in parallel with resistive and capacitive components, with the total circulating current being: 
\begin{equation}
    I(t) = I_c \sin{\phi} + \hbar G\dot{\phi} + \hbar C \ddot{\phi}.
    \label{Eq:RCSJmodel}
\end{equation}
The supercurrent contribution is given by the sinusoidal current-phase relation $I_c \sin{\phi}$, the resistive and capacitive contributions are $G \dot{\phi}$ and $C \ddot{\phi}$, respectively.
Similarly as for SJJ, the Josephson-Anderson relation $\Delta \mu = -\hbar \dot \phi$ connects the phase dynamics to the chemical potential build-up \cite{Smerzi1997, Meier2001}, with  $\Delta \mu$ playing the role of the junction voltage. 
The Stewart-McCumber parameter $\beta_c = I_c C/ \hbar G^2$ determines the dynamic regime of the junction, discriminating between underdamped ($\beta_c \gg 1$) and overdamped ($\beta_c \ll 1$) 
\cite{Tinkham2004}. 
In the latter, the undriven solution is analytically given by $\Delta \mu = G^{-1}\sqrt{I_{DC}^2-I_c^2}$, 
which we use to fit the DC measurements
to extract $I_c$ and $G$. 
In the driven scenario instead, the $I-\Delta \mu$ characteristic is evaluated by numerically solving Eq.\ref{Eq:RCSJmodel}. As reported in Fig.~\ref{fig2} A-B,
the overdamped solutions of the RCSJ model display the characteristic Shapiro steps matching the measured ones. 
We note that even though we 
estimate  
$\beta_c\approx 14(2)$ for the unitary superfluid, the overdamped RCSJ numerical solutions well capture the observed steps.
%
In particular, the step widths obtained from the numerical solutions of the overdamped RCSJ model are in agreement with the experimentally measured values (Fig.~\ref{fig2} D). Note that this result is analogous to the Bessel function behavior of the Shapiro steps found analytically under voltage-driven modulation \cite{Barone1982, Tinkham2004, Simmonds2001, SM}, $\Delta I_n/I_c = \left|J_n \left(V_{ac}/\hbar \omega\right)\right|$, but quantitatively deviates from it at small driving amplitudes, as expected under current-driven modulation \cite{Panghotra2020}.  


\begin{figure*}[ht!]
\centering
\includegraphics[width=\textwidth]
{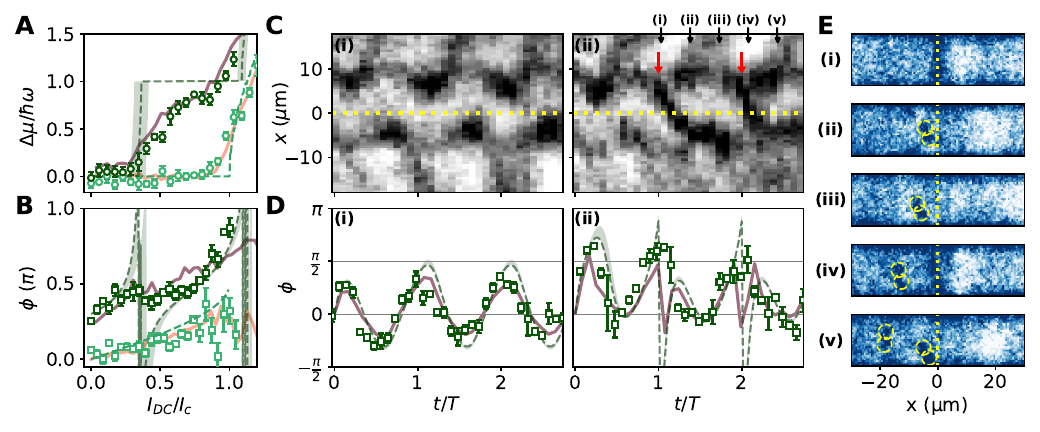}
\caption{\textbf{Phase dynamics and dynamical locking in a driven atomic JJ.}
(\textbf{A}) Current-voltage characteristic for a BEC junction under DC (light green symbols), and DC+AC drive at $\omega = 2 \pi \times 175\, $Hz and $I_{AC}/I_{C} = 1.4(1)$ (dark green symbols). 
(\textbf{B}) Current-phase relation under same experimental conditions as \textbf{A}, vertically shifted by $0.3\, \pi$ for the DC+AC case. 
Dashed lines represent the fit (light green) with the analytical and numerical solutions (dark green) of the overdamped RCSJ model. The shaded area accounts for the $I_c$ fitting error.
(\textbf{C}) Time evolution of the interference pattern between the reservoirs for $I_{AC}/I_{c} = 1.4(1)$ and $I_{DC}/I_c = 0.14$ (\textbf{i}), $I_{DC}/I_c = 0.83$ (\textbf{ii}).
Each row reports the integrated fringe profile, averaged over 5 repetitions.
Red arrows signal the occurrence of phase slips; the yellow dotted line marks the barrier position.
(\textbf{D}) Measured phase time-evolution from a sinusoidal fit of the fringe patterns.
Solid lines in panels (A-B-D) represent the numerical simulation results 
(see main text), with $I_{AC} = 0$ (orange) and $I_{AC}/I_c = 1.4$ (dark purple). 
(\textbf{E}) Density profile in time-of-flight at $t/T= 1.03 $ (\textbf{i}), $1.38$ (\textbf{ii}), $1.73$ (\textbf{iii}), $2.08$ (\textbf{iv}), $2.43$ (\textbf{v}). Vortices are marked in yellow dashed circles.
\label{fig3}}
\end{figure*}

\section*{Phase dynamics and synchronization}

To link the 
Shapiro steps with the phase dynamics, we tune our atomic junction to operate in the BEC regime at $1/k_Fa=3.3(1)$. Here, 
the reduced interactions 
allow to leverage matter-wave interference between the expanding reservoirs to directly measure the relative phase $\phi$ \cite{Li2024}, accessing 
both $\Delta\mu-I$ and $\phi-I$ characteristics \cite{Kwon2020}, as reported in Fig.~\ref{fig3} A-B 
for $\omega = 2 \pi \times \SI{175}{Hz}$, $I_{AC}/I_c = 0$, 
and $I_{AC}/I_c =1.4(1)$. 
In the non-modulated case for $I_{DC}<I_c$, we recover the expected DC Josephson trend \cite{Kwon2020}, \textit{i.e.} $\phi = \arcsin(I_{DC}/I_c)$.
The measured $\phi$ shows larger errorbars only for currents near and above $I_c$, attributed to the running phase in the voltage state \cite{SM}.
%
%
%
%
The driven scenario instead displays two main features: a diminished step contrast  in the $I-\Delta \mu$ curve as compared to the unitary junction, and a well-defined phase $\phi$ 
within the plateau region even at finite $\Delta \mu$, presenting significant fluctuations only in the transition regions between plateaus \cite{SM}.
To support our findings, 
we compare the experimental results 
with numerical simulation performed with a dynamical classical-field method under similar experimental conditions \cite{SM, Singh2016}.
The numerical simulations quantitatively agree with our results for $\Delta\mu-I$ and $\phi-I$ characteristics, capturing the smoothness of the Shapiro steps in the experimental data.
The agreement with the numerical simulation evidences that the observed lower contrast of the plateaus in the BEC is inherent to our junction, which nevertheless displays Shapiro steps dynamics in a broad range of modulation parameters
(Fig.~\ref{figSM_BEC_Shap} of Supplementary Materials).

We demonstrate the dynamical synchronization between the junction phase and the external driving by monitoring $\phi$ as a function of time 
in the 0-th and 1-st Shapiro steps (Fig.~\ref{fig3} C), additionally, the 2-nd Shapiro step is shown in Fig.~\ref{fig:sup2ndStepPhase} of Supplementary Materials.
The interference fringes show an oscillating behavior in both cases, highlighted by 
the measured relative phase 
$\phi$ (Fig.~\ref{fig3} D), which showcases the phase-locking and synchronized dynamics induced by the driving.
%
%
%
%
Note that %
phase locking effects have also been observed in  a superfluid system in a Floquet driven tilted double well potential \cite{Ji2022}.
Whereas in the $0$-th step, the phase oscillates in phase with the injected current without reaching 
the critical values for phase slips, in the $1$-st step $\phi$ reaches this critical limit once every cycle, except for the transient behavior of the first modulation, leading to periodic 
phase slips, marked in Fig.~\ref{fig3} C-D (ii) by the red arrows.
The phase-slippage process produces a clear spike in $\phi$ for both the dynamical classical-field simulation and overdamped RCSJ solution, associated to an increase of the errorbar on the measured phase \cite{SM, Burchianti2018}.

Depending on the 
junction dimensionality, phase slips manifest as different topological excitations, from a soliton in 1D
\cite{Tinkham2004, Binanti2021}, to vortex-antivortex pair in 2D \cite{Singh2023}, and vortex rings in 3D \cite{Burchianti2018}. 
The kind of excitations created is determined by which is energetically favorable \cite{VanAlphen2019}, or arises from 
a multi-step process such as soliton decay into different topological excitations
\cite{VanAlphen2019, Ku2016}.
In our BEC junction, we directly observe the presence of vortex-antivortex pairs as
localized density depletions in the superfluid after a short time of flight \cite{SM} (Fig.~\ref{fig3} E). 
For the $1$-st Shapiro step, we observe vortices generated on the near-left of the barrier (i-ii) that subsequently move towards the bulk of the reservoir (iii-iv). After an additional modulation period, a second vortex-antivortex pair is similarly emitted from the barrier region (v).
The dynamical classical-field simulations corroborate this mechanism (Fig.~\ref{supFigVJsimMovie}-\ref{supFigVJsimMovie_c2} of Supplementary Materials): 
at the end of the first cycle, the 
phase difference accumulated along the junction releases a solitonic excitation from the barrier, which 
then quickly decays into vortex-antivortex pairs, via snake-instability \cite{SM, VanAlphen2019}.
The periodic emission of vortex-antivortex pairs in Shapiro steps is a proxy of the underlying synchronized phase dynamics of the junction, which undergo $n$ phase slippage processes in the $n$-th step of each modulation cycle, as pictorially illustrated in the washboard potential analogy of Fig.~\ref{fig1} C(ii) for $n=1$. 

\begin{figure}[t!]
\centering
\includegraphics[width=5.7cm]{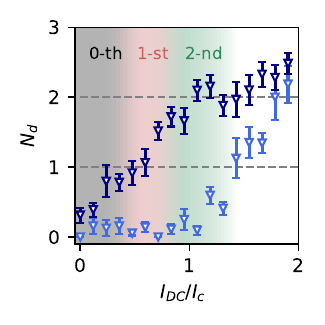}
\caption{\textbf{Vortex-antivortex pairs as phase-slips proxis in a unitary JJ.}
Number of emitted vortex-antivortex pairs as a function of the bias current $I_{DC}/I_c$ for $I_{AC}/I_c = 0$ (light blue), and $I_{AC}/I_c = 2.8(2)$ (dark blue). We measure the number of vortices after three periods of driving with a frequency $\omega=2\pi \times 175$ Hz. The background color highlights the regions of the $0$-th, $1$-st, and $2$-nd Shapiro steps, with a color gradient reflecting the measured $\Delta \mu$ under the same driving condition. Error bars represent the standard deviation of the mean over 3 to 5 realizations.
}
\label{fig4}
\end{figure}

By measuring the number of vortex pairs $N_d$ as a function of the injected current (Fig.~\ref{fig4}), we probe the synchronization also in the strongly-interacting regime.
In the non-modulated case
, vortices are observed to proliferate for $I_{DC}>I_c$, signaling a contribution to the junction conductance due to collective excitations, in agreement with previous observations in three-dimensional atomic JJs \cite{DelPace2021,Burchianti2018}.
For the AC drive instead
, the number of detected vortex pairs shows a step-like trend that closely resembles the Shapiro steps observed in the $I- \Delta \mu$ curve measured under the same driving conditions. 
In particular, in the $2$-nd Shapiro step, $N_d$ doubles with respect to the $1$-st step, highlighting that, as a result of the underlying synchronized phase dynamics, the number of phase slips is proportional to the step number. 
%
%
The dissipative and resistive dynamics giving rise to the Shapiro steps in our atomic JJ is dominated by collective excitations, playing a similar role of quasiparticles produced by Cooper pair breaking in SJJ. In fact, in all explored configurations, the junction always works in the regime of  $\Delta \mu \ll \Delta$, with $\Delta$ the superfluid gap, where broken pairs are energetically suppressed and the only accessible excitations are collective ones, namely phonons, solitons and vortices \cite{DelPace2021}. 

\section*{Conclusions}

We have observed Shapiro steps in a current driven atomic JJ composed of two weakly-coupled strongly-interacting Fermi gases. 
Shapiro steps appear as quantized plateaus in the relative chemical potential, the analog of the voltage drop in SJJs, occurring at integer multiples of the external driving frequency. As the driving amplitude increases, the widths of these plateaus exhibit a non-linear dependence, arising 
from the 
interplay between the external drive and the non-linear characteristics  of the atomic Josephson junction.  
We probe the synchronization process at the heart of Shapiro steps by directly accessing the microscopic phase dynamics of the junction featuring periodic phase-slippage processes 
which manifest as vortex-antivortex pairs. 
At the more fundamental level, our results also disclose the interplay between quantum phase coherence and dissipation dynamics in a strongly correlated Fermi system.
Our work opens prospects for a deeper comprehension of how quasiparticles and quantum phase slips contribute to quantum transport, potentially leading to advancements in understanding the complex dynamics within driven superconducting networks \cite{guichard2010phase,arutyunov2008superconductivity,pop2010measurement}. 
Our driven atomtronic circuit 
can be indeed exploited
to investigate driven atomic JJ arrays, whether arranged linearly or in annular configurations \cite{Pezz2024}. In turn, this opens the way to simulate important paradigms in quantum synchronization, as provided by Kuramoto-like models \cite{Kuramoto1984}, by driven atomic JJ in the presence of strong interactions and with disordered couplings between the links \cite{Watts1998, Trees2005}.
Finally, as Shapiro steps provide the voltage standards in quantum electronics \cite{Kohlmann2011}, we could leverage the quantized steps to access a direct measurement of chemical potential differences \cite{Kohler2003}. 
This approach would be specifically  valuable in the intermediate regimes of the crossover superfluids, where the equation of state is challenging  \cite{MakingProbingFermi}.



We note that Shapiro steps have been very recently observed also in a driven atomic JJ with weakly interacting bosons \cite{Bernhart2024}.




\section*{Acknowledgments}
We thank Francesco Marino, Ludwig Mathey, and Klejdja Xhani for the discussions and Francesco Scazza for careful reading of the manuscript. G.R. and G.D.P. acknowledge financial support from the PNRR MUR project PE0000023-NQSTI. G.R. acknowledges funding from the Italian Ministry of University and Research under the PRIN2017 project CEnTraL and the Project CNR-FOE-LENS-2023. The Authors acknowledge support from the European Union - NextGenerationEU for the “Integrated Infrastructure initiative in Photonics and Quantum Sciences" - I-PHOQS [IR0000016, ID D2B8D520, CUP B53C22001750006]. This publication has received funding under the Horizon Europe program HORIZON-CL4-2022-QUANTUM-02-SGA via project 101113690 (PASQuanS2.1).
J.A.S. acknowledge financial support from CONAHCyT grant CF-2023-I-72; DGAPA-UNAM-PAPIIT grant IN105724; CIC-UNAM grant LANMAC-2024, and European Community's Horizon 2020 research and innovation program under grant agreement n° 871124.

\section*{Author contributions}

G. D. P., V. P. S., L. A. and G. R. conceived the experiment. 
G. D. P., D. H.-R., N. G., M. F. F. and G. N. performed the experimental work, acquired the data and carried out the data analysis. 
V. P. S. performed numerical simulations. 
L. A and G. R. supervised the work.
All authors contributed extensively to the discussion and interpretation of the results, and took part in writing the manuscript.



\bibliography{scibib}
\bibliographystyle{Science}

\renewcommand{\thefigure}{S.\arabic{figure}}
\setcounter{figure}{0}
\renewcommand{\theequation}{S.\arabic{equation}}
\setcounter{equation}{0}
\renewcommand{\thesection}{S.\arabic{section}}
\setcounter{section}{0}
\renewcommand{\thetable}{S.\arabic{table}}
\setcounter{table}{0}
\clearpage
\section*{Supplementary materials} 

\subsection*{Experimental method}
\subsubsection*{Gas preparation}
We initially realize a superfluid atomic Fermi gas by evaporatively cooling the first and third lowest hyperfine state of $^6\text{Li}$ atoms in a red-detuned, cigar-shaped harmonic trap at $\SI{690}{G}$. To obtain a superfluid in the BEC regime, during the last part of the evaporative cooling procedure we sweep the magnetic field at $\SI{633}{G}$, where the molecular scattering length is $a_M = 0.6\, a = 1029\, a_0$. After the production of the superfluid, we adiabatically ramp up in $\SI{100}{ms}$ a TEM$_{0,1}$-like laser beam, that confines the atoms in the $\hat{z}$ direction creating a harmonic potential, and the hard wall potential with a rectangular shape in the $x-y$ plane created with the DMD. Both these two potentials are realized with a blue-detuned laser at $\SI{532}{nm}$. The vertical harmonic confinement has a trap frequency $\omega_z =\mathrm{2\pi\times}\SI{685(5)}{Hz}$ at unitarity and $\omega_z =\mathrm{2\pi\times}\SI{416(4)}{Hz}$ in the BEC regime. Subsequently, we adiabatically ramp down the initial cigar-shape potential in $\SI{100}{ms}$, and we let the system equilibrate for at least $\SI{50}{ms}$ before ramping up the tunneling barrier. 
The superfluid produced in such a hybrid trap is homogeneous in the $x-y$ plane since the residual harmonic confinement arising from the TEM$_{0,1}$-like and the curvature of the magnetic field corresponds to $\SI{2.5}{Hz}$, therefore negligible for the dynamics studied in this work. 
The Fermi energy of the system in the hybrid trap is computed directly from a Thomas-Fermi approximation yielding \cite{DelPace2022}: $E_F = \hbar^2 k_F^2/2m = 2 \hbar \sqrt{\hbar \pi \omega_z N/m A}$, where $\hbar$ is the reduced Plank constant, $m$ the mass of a $^6$Li atom, and $A = L_x \times L_y$, the area of the box trap. 
The chemical potential of the unitary superfluid is calculated as $\mu_0/h = \xi^{3/4} E_F/h = 5.9 (2) \, $kHz, where $\xi \simeq 0.37$ is the Bertsch parameter; whereas, for the molecular BEC, we calculate it as $\mu_0/h = (3 \pi N \hbar^2 \omega_z a_M / 2 A \sqrt{m} )^{2/3} / h = 1.20(5) \,$kHz \cite{DelPace2021}.
The healing length in the BEC regime is $\xi_L = \SI{0.59(1)}{\mu m}$, while in the UFG the typical length scale of the superfluid is given by $2\pi/k_F \simeq {1.67(3)}{\,\mu m}$. 
To produce the atomic JJ, we ramp up the barrier in its initial position in $\SI{10}{ms}$ and wait for $\SI{30}{ms}$ before starting the movement. The homogeneity of the cloud before ramping up the barrier ensures the same density and chemical potential in the two reservoirs at $t = \SI{0}{ms}$.

\begin{figure}[t!]
\centering
\includegraphics[width=0.45
\textwidth]{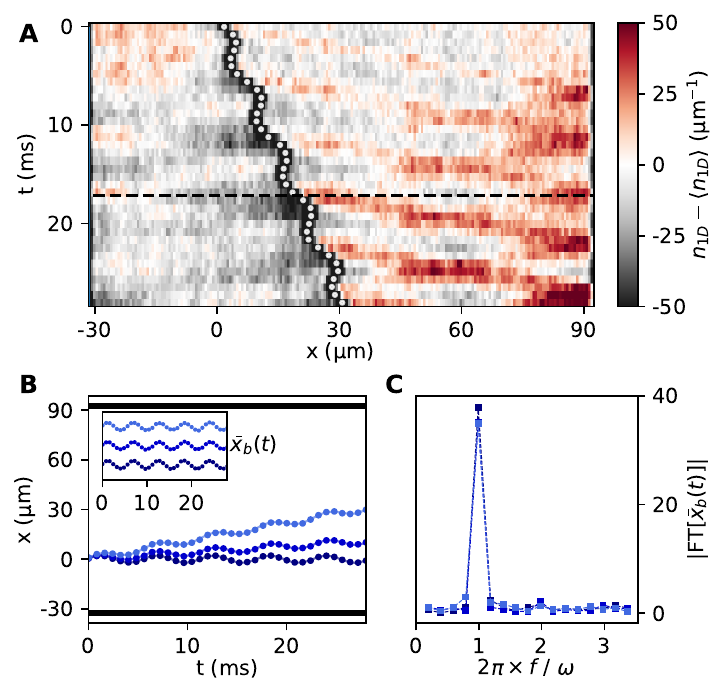}
\caption{\textbf{Density time evolution} (\textbf{A}) Time evolution of the relative integrated density along the $y$ direction, $n_{1D}$, with respect to the average value at each time $\langle n_{1D} \rangle$ for $v_{DC} = \SI{1.1}{\mu m/ms}$, $\omega = \mathrm{2\pi\times}\SI{175}{Hz}$ and $x_{AC} = \SI{2.0}{\mu m}$. Density modulations arising from the instantaneous chemical potential difference at the barrier position $\Delta \mu_b (t)$ travel inside the two reservoirs. The horizontal black dashed line indicates the time at which the measurement is performed. White points show the extracted position of the barrier as a function of time. 
(\textbf{B}) Trajectories of the barrier for $\omega = \mathrm{2\pi\times}\SI{175}{Hz}$, $x_{AC} = \SI{2.0}{\mu m}$ and $v_{DC} = \SI{0}{\mu m/ms}$ (dark blue points), $v_{DC} = \SI{0.4}{\mu m/ms}$ (medium blue points) and $v_{DC} = \SI{1.1}{\mu m/ms}$ (light blue points). 
The black solid lines markes the edges of the junction.
The inset shows the barrier position with respect to the trajectories with the same parameters as extracted by the fits, but $x_{AC} = 0$. Curves corresponding to different velocities are shifted for better visualization. (\textbf{C}) Fourier transform of the quantity in the inset of \textbf{B}. Dashed lines are guides to the eye.}
\label{figSM_density_evolution}
\end{figure}

The chemical potential difference $\Delta \mu$ 
is extracted by taking an \textit{in situ} image of the cloud after the barrier movement and  
computing $\mu_L$ ($\mu_R$) from the measured number of atoms $N_L$ ($N_R$) in the left (right) reservoir. This way of computing $\Delta \mu$ is analogous to a measurement of the time average $\langle \Delta \mu \rangle_t$. In fact, as it is shown in Fig.~\ref{figSM_density_evolution} A, the instantaneous density accumulation and depletion, created in the vicinity of the barrier because of its motion, propagate in time away from the barrier position inside the two reservoirs, traveling at the speed of sound (measured as in Ref. \cite{HernndezRajkov2024}). 
Consequently, the time evolution of the instantaneous chemical potential difference at the barrier position, $\Delta \mu_b (t)$, is mapped in different positions in space. 
The black horizontal dashed line in the figure indicates the time at which the measurement is performed, corresponding to three modulation periods. At this time, the initial density modulations have traveled up to the edge of the cloud covering the full reservoirs. For this reason, the time average value of $\Delta \mu_b (t)$ is mapped in the measured value of the global chemical potential difference, $\Delta \mu$, between the two reservoirs.
This reasoning is expected to be less valid for higher frequencies, for which the period is shorter and the density modulations do not reach the edges at the end of the three cycles, leading to a 
lower value of the measured $\Delta \mu$. 
Given the lower speed of sound in the BEC regime, the decrease happens at lower frequencies with respect to the UFG regime. 
We note that all the measurements of $\Delta \mu$ reported in this work have been performed by imaging the cloud after three modulation periods. This value is the longest number of cycles allowed by the size of the junction, for the range of barrier velocity spanned. For each $\omega$, we also acquired a corresponding DC curve with $I_{AC}=0$, moving the barrier at constant velocity for a time of $3 \,T$.

\subsubsection*{Barrier characterization}
The barrier intensity, shape, and position are controlled with the DMD. The effective movement of the barrier is obtained by creating a sequence of different light patterns with the DMD, with the barrier position following the desired equation of motion 
and shining it on the atomic plane through the high-resolution microscope objective with a frame rate of $12.5 \,$kHz. 
Even though the DMD-made barrier movement is intrinsically discretized, the small size of a single DMD-pixel imaged on the atomic plane, $0.25 \, \mu$m, ensures the smoothness of the barrier movement. 
We characterize the properties of the barrier movement by monitoring the position of the barrier as a function of time in the case of Fig.~\ref{figSM_density_evolution} B. The barrier position is obtained by performing a Gaussian fit of the density depletion of the \textit{in situ} images as a function of time. The extracted positions are indicated by the white points in Fig.~\ref{figSM_density_evolution} A. We fit the trajectory with the equation of motion of the barrier $x(t) = v_{DC} t + x_{AC} \sin (\omega t)$. The fitted parameters are compatible within $\SI{1}{\%}$ error for the value of $v_{DC}$ and $\SI{0.5}{\%}$ error in the case of $x_{AC}$ and $\omega$ with respect to the set values.
The Fourier transform (FT) of the observed trajectories respect the ones with the same parameters but $x_{AC} = \SI{0}{\mu m}$, $\bar{x}_b(t)$, shows an almost single-frequency behavior, with a small contribution of the second and third harmonics, as it is shown in Fig.~\ref{figSM_density_evolution} C, confirming that our protocol provides a clean and monochromatic AC drive. The measurements reported in this work have been performed with values of $x_{AC}$ in the range $1-6 \, \mu$m, limited by the finite resolution of the DMD-projecting setup and the frame rate of the DMD, respectively.

\begin{figure}[ht!]
\centering
\includegraphics[width=0.45
\textwidth]{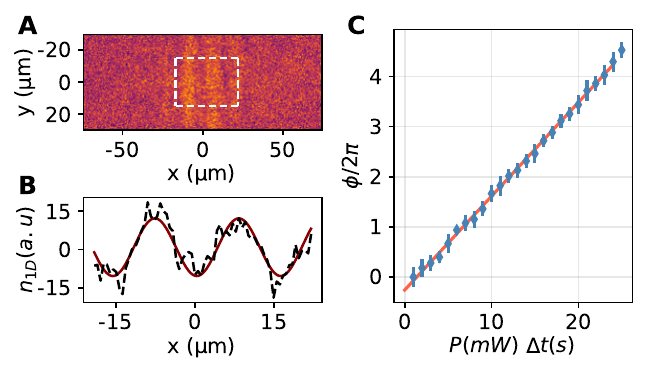}
\caption{\textbf{Phase imprinting calibration technique} (\textbf{A}) Example of an interference pattern between the two sides of the JJ after the phase imprinting (\textbf{B}) 1D integrated density profile of the interference pattern for the selected region delimited by the white dash lines. The sinusoidal fit allows for the extraction of the relative phase between the imprinted and non-imprinted reservoir (\textbf{C}) Relative phase for different powers of the imprinting light pulse, considering a fixed imprinting time of $\Delta t = \SI{500}{\mu s}$. With a linear fit we obtain the calibration parameter $\alpha= \SI{0.187(1)}{Hz/mW}$.}
\label{figS_calibration}
\end{figure}

The intensity of the profile created to produce the tunneling barrier is calibrated with a phase imprinting method \cite{Kwon2020, Luick2020}. 
We create a small JJ of dimension $37.5 \times 17.5 \, \mu$m$^2$ and imprint a phase on one of its reservoir
by applying a homogeneous optical potential $U_0$ for a short time interval $\Delta t$. For $\Delta t < h/\mu$, we operate an 
almost pure phase excitation, with the phase imprinted by the light on the illuminated reservoir given by $\hbar \phi = U_0 \Delta t$. To calibrate the potential $U_0$, we measure $ \phi$ from the interference arising in the time-of-flight (TOF) expansion of the junction, 
using the phase of the non-imprinted reservoir as a reference. 
Figure \ref{figS_calibration} A-B shows an example of an interference pattern, which we fit to extract $\phi$ with a sinusoidal function. 
By repeating this process for different values of the imprinting pulse power, we extract a calibration for the imprinted potential $U_0$ (Fig.~\ref{figS_calibration} C).

To characterize the barrier height and width we image the DMD-produced barrier intensity pattern on a service camera in the DMD-projecting optical system. We acquire images of the barrier in different positions throughout the area of the JJ and extract their height and size via a Gaussian fit along the $ x-$ direction. To account for the finite resolution of the imaging system after the service camera, before performing the fit, we convolve the images with a Gaussian of FWHM$= 0.63 \, \mu$m, well-approximating the Point Spread Function (PSF) of the microscope objective. 
The barrier properties are barely varying while changing their position, providing the average values reported in the main text for $V_0$ and FWHM. In particular, to obtain the value of the barrier height $V_0$ we compare the fitted barrier height from the image with the pattern used to create the phase imprinting and employ the calibration factor to convert it into energy units. 

\subsubsection*{Phase and vortex detection}
\begin{figure}[ht!]
\centering
\includegraphics[width=0.45\textwidth]{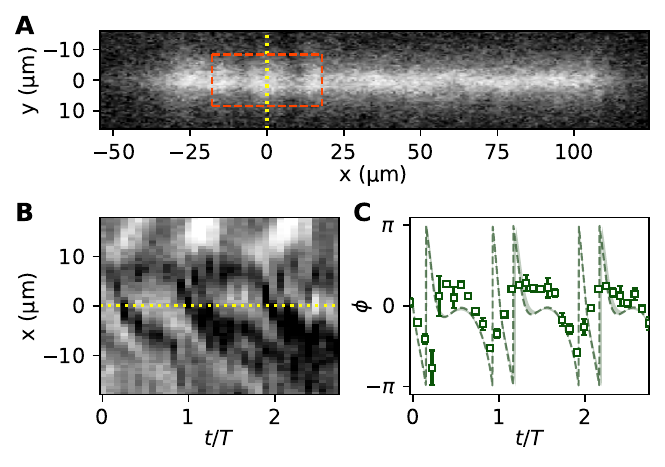}
\caption{\textbf{Phase evolution in the 2nd Shapiro step}
(\textbf{A}) The interference pattern between the reservoirs for the specific time $t_0 = 0$ ms of the DC+AC driving of the barrier. The red rectangle represents the region used for obtaining the 1D mean density profile that is fitted to measure the phase. The yellow dotted line marks the position of the barrier. (\textbf{B}) Interference pattern as a function of time for $I_{DC}/I_c = 1.8(1)$ and $I_{AC}/I_c = 1.4(1)$. 
Each row corresponds to the integrated fringe profile along the y direction, averaged over 5 repetitions. (\textbf{C}) Extracted phase from the interference fringe pattern as in B via a sinusoidal fit of the fringe contrast. The dashed line represents the RCSJ model result for the driven situation.
}
\label{fig:sup2ndStepPhase}
\end{figure}

The relative phase between the two reservoirs is measured from the interference fringes arising after a TOF expansion of $2\,$ms, after abruptly switching off all the confinements. The short TOF employed for this measurement ensures that interference fringes appear only close to the barrier region (Fig. \ref{fig:sup2ndStepPhase}), allowing for a precise measurement of the relative phase at the junction $\phi$. In particular, to quantitatively extract $\phi$ we restrict to a region around the instantaneous position of the barrier (dashed red rectangle area in the figure) and fit the density profile integrated along the $y$-direction with a sinusoidal function. We constrain the wavelength of the fringes to the same value for all the acquired interferograms, as this quantity depends only on the TOF. We fix the best fringe wavelength as the average wavelength from a preliminary sinusoidal fit to all the data acquired under different conditions. In Fig. \ref{fig:sup2ndStepPhase} B-C we report the measured phase evolution in the second Shapiro step, under the same conditions as in Fig.~\ref{fig3} of the main text. 
Both in the experimental data and in the overdamped RCSJ model numerical solutions, two phase-slips processes are evident in each modulation cycle. 

To detect vortices in the BEC regime we employ a TOF technique as well, but, to avoid the simultaneous presence in the images of interference fringes, we ramp down the barrier in $\SI{0.24}{ms}$, wait $\SI{4}{ms}$ and then perform the TOF expansion. To increase the contrast of the vortices, we ramp down the DMD-made potential during the first $1\,$ms of TOF expansion. 
To observe the presence of vortices in the UFG regime we add to the TOF expansion a fast sweep of the Feshbach magnetic field to map the system in a molecular BEC, where vortices are visible as clear holes in the density \cite{DelPace2022}.
In particular, at the end of the movement of the barrier
, we wait $\SI{1}{ms}$ 
before rapidly removing the barrier from the system in $\SI{1.5}{ms}$ 
and then we wait $\SI{1.5}{ms}$ 
before performing the TOF procedure, 
starting the magnetic field ramp 
after the end of the barrier movement.
In both BEC and UFG regimes, the waiting time between the barrier ramp down and the TOF expansion is expected not to affect significantly the number of vortices in the system. 
We note that, since $N_d$ of Fig.~\ref{fig4} of the main text is measured after three modulation cycles, we would expect to see a higher number of vortex-antivortex pairs, corresponding to the periodic phase-slippage process. 
However, the finite size of the junction along the $y$-direction facilitates the vortices to escape from the bulk density, so that the measured $N_d$ most likely corresponds to the number of vortex-antivortex pairs emitted during the last modulation cycle.
We note that with a similar protocol as for detecting vortices, employing a fast sweep during the TOF, the relative phase at the junction could be measurable also at unitarity from the interferograms \cite{DelPace2022}. However, this technique provides more noisy interferograms, and the lower signal/noise together with the presence of vortices makes the extraction of $\phi$ harder. For this reason, we decided to perform the phase measurement in the BEC regime.

\subsection*{Circuital model simulations}

We study the expected behavior of $\Delta \mu$ and of $\phi$ in the framework of the RCSJ model, in which the phase evolution is described by Eq.~(\ref{Eq:RCSJmodel}) of the main text. We solve numerically the equation in the overdamped regime ($\beta_c \ll 1$), neglecting the capacitance, i.e. the second derivative term. 
In this limit, the $I_{AC}=0$ case solution $\Delta \mu = G^{-1}\sqrt{I_{DC}^2-I_c^2}$ is used to extract the values of $G$ and $I_c$ by fitting the observed value of $\Delta \mu$ as a function of $I_{DC}$. These values are then used in order to simulate the dynamic in the $I_{AC}\neq 0$ case. Figure \ref{figS2} displays the results for values of the parameters as the case plotted in Fig.~\ref{fig3}A-B of the main text, using the Josephson-Anderson relation $\Delta\mu = -\hbar \dot{\phi}$. In particular, we extract the expected average value of $\Delta\mu$ from the time average value of the phase derivative $\langle \dot{\phi} \rangle$ for a total evolution time of $\SI{10}{periods}$. Figure \ref{figS2}C shows the expected behavior in the absence of modulation ($I_{AC}=0$, blue line) and the predicted shape of Shapiro steps for a $I_{AC}=1.3 \,I_c$. In Fig.~\ref{figS2}D we plot the corresponding phase at the end of 3 complete periods for the same cases of $I_{AC}=0$ and $I_{AC}=1.3 \,I_c$. This plot shows a similar trend to the one shown in Fig.~\ref{fig3}B of the main text, apart for the regions in the transition between the steps in the modulated case, where this quantity shows a not well-defined behavior. In particular, Fig.~\ref{figS2}E, shows the value of the distance between the phase at a given value of $I_{DC}$ resulting from different initial values of the phase $\phi_0$: $\Delta \phi^2 (I_{DC}) = \sum_{i,j} ( \phi(I_{DC},\phi_0 = \phi_{0,i}) - \phi(I_{DC},\phi_0 = \phi_{0,j}))^2$. This value diverges in this transition region.

\begin{figure}[ht!]
\centering
\includegraphics[width=0.48\textwidth]{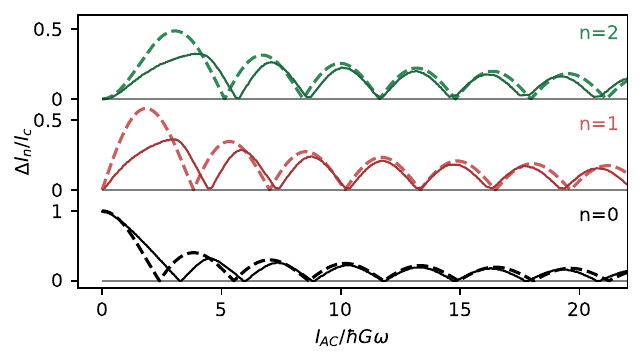}
\caption{\textbf{Width of Shapiro steps under current modulation.} 
Current modulated half-widths obtained from the overdamped RCSJ simulations are shown as continuous curves. Dashed curves correspond to the Bessel behavior $\Delta I_n/I_c = |J_n(V_{AC}/\hbar \omega)|$ for voltage modulation.
}
\label{figS1b}
\end{figure}

\begin{figure*}[ht!]
\centering
\includegraphics[width=0.8
\textwidth]{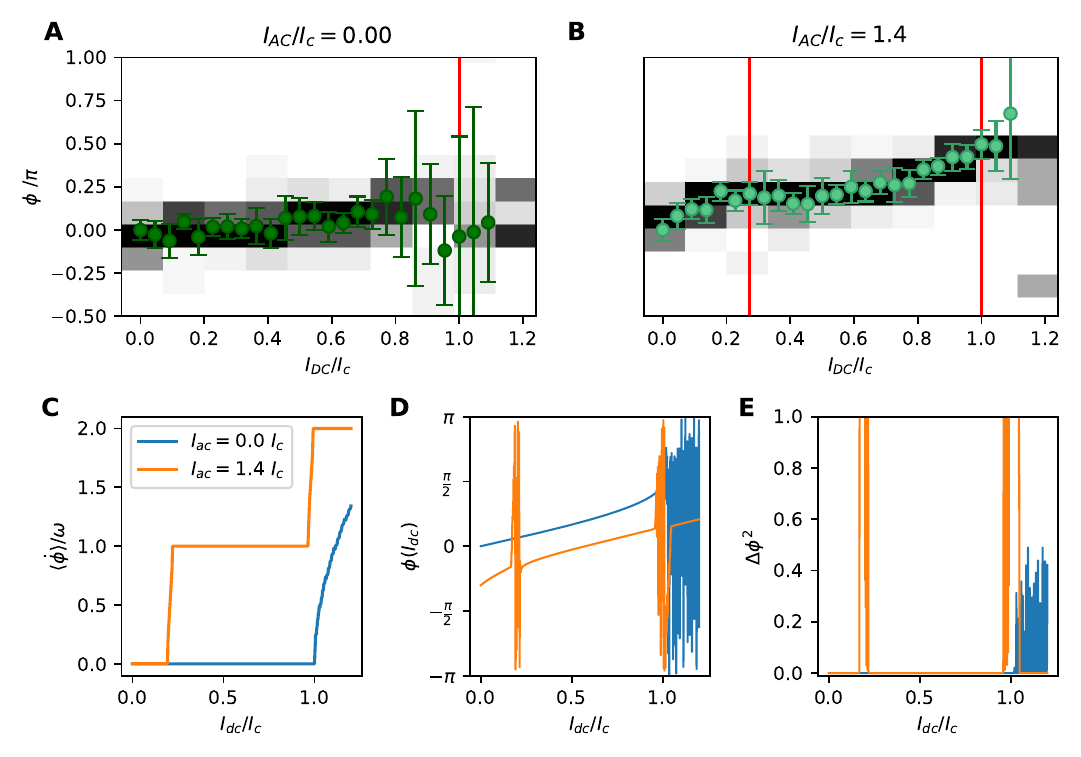}
\caption{\textbf{Phase fluctuations in the $\phi-I$ characteristics}
Current-phase characteristic for a BEC junction for a (\textbf{A}) DC, and DC+AC drive at $\omega=2\pi\times 175$ Hz and (\textbf{B}) $I_{AC} /I_c = 1.4(1)$ (green symbols), obtained from RCSJ circuit model solution with different initial phase conditions. The background shows the histogram of measured phases as a function of $I_{DC} /I_c$. Red vertical lines denote the critical current and the transition from the 0-th to the 1-st Shapiro step at $I_{DC} /I_c = 0.27(1)$.
(\textbf{C}) Time average $\langle \dot \phi \rangle$, (\textbf{D}) average phase 
and (\textbf{E}) Variance of $\phi$ measured after three modulations over 100 different initial conditions in the domain $[-2\pi/10, 2\pi/10]$, for $I_{AC} /I_c = 0$ (blue) and $1.4(1)$ (orange).
}
\label{figS2}
\end{figure*}

Under a voltage modulation, the Shapiro step half-widths follow the well-known relation $\Delta I_n/I_c = |J_n(V_{AC}/\hbar \omega)|$. However, quantitative deviations occur under current modulation, especially for the first few steps, as shown in Fig.~\ref{figS1b}.

\subsection*{Phase Fluctuations}

In our system, we refer to the phase fluctuations in terms of the fluctuations of the relative phase given small deviations in the initial conditions of $\phi$, and not to stochastic noise terms in the RCSJ model.
These fluctuations in the initial conditions are natural in our experimental setup and modify the measurement of the phase we perform after TOF, as discussed in the previous section.

To quantify the stability of the phase measurements we make use of the overdamped RCSJ model. We compare the phase evolution of from 100 simulations of Eq.~\eqref{Eq:RCSJmodel} given different initial conditions $\phi(t=0) = \varepsilon$, where $\varepsilon$ follow a uniform distribution in the domain $[-2\pi/10, 2\pi/10]$. The $\varepsilon$ range is chosen based on the fluctuations observed without any external current injected in the JJ. We compare our experimental data to the simulation results after three modulation cycles, see Fig.~\ref{figS2}.
Let us note that the data points in Fig.~\ref{figS2} A-B display a larger error bar near the transitions between the plateau regions (B), and near and above $I_c$ (A), also visible in the measured $\phi$ as larger errorbars. 
This behavior is recovered from the numerical simulations shown in Fig.~\ref{figS2} D-E, where the variance measured over the ensemble of different initial conditions is non-zero only in the transition regions while maintaining a well-defined value in the central region of the plateaus.

\subsection*{Classical-field simulation method}
To simulate the dynamics of $^{6}\text{Li}_2$ molecules on the BEC side we use classical-field dynamics within the truncated Wigner approximation \cite{Singh2016,Singh2020}.
The condensate is described by the Hamiltonian
\begin{widetext}
\begin{align} \label{eq:hamil}
\hat{H} = \int \mathrm{d}{\bf r} \Big[  \frac{\hbar^2}{2m}  \nabla \hat{\psi}^\dagger({\bf r}) \cdot \nabla \hat{\psi}({\bf r})  + V({\bf r}) \hat{\psi}^\dagger({\bf r})\hat{\psi}({\bf r})   + \frac{g}{2} \hat{\psi}^\dagger({\bf r})\hat{\psi}^\dagger({\bf r})\hat{\psi}({\bf r})\hat{\psi}({\bf r})\Big],
\end{align}
\end{widetext}
where $\hat{\psi}({\bf r})$ and $\hat{\psi}^\dagger({\bf r})$ are the bosonic annihilation and creation field operators, respectively. 
The 3D interaction parameter is given by $g=4\pi a_D \hbar^2/m_{D}$, where $a_D$ is the molecular $s$-wave scattering length and $m_{D} = 2 m$ is the molecular mass.
Following the experiments we choose: $a_D = 0.6 a=1029\, a_0$, where $a_0$ is the Bohr radius, $V({\bf r}) = V(z) = m_D \omega_z^2 z^2/2$ the harmonic trapping potential, with $\omega_z = 2 \pi \times 416 \, $Hz being the trap frequency in the $z$ direction. 
%
For our simulations, we consider the same dimensions of the experimental Josephson junction, by mapping the real space on a lattice system of  $253 \times 37 \times 8$ sites with the lattice discretization length $l= 0.5\, \mu \,$mm.
We note that $l$  is chosen such that it is to be comparable or smaller than the healing length $\xi_L = \hbar/\sqrt{2mgn_0}$ and the de Broglie wavelength, where $n_0$ is the density \cite{Mora2003}. 
In the classical-field representation, we replace the operators $\hat{\psi}$ in Eq. \ref{eq:hamil} and in the equations of motion by complex numbers $\psi$. 
We sample the initial states $\psi(t=0)$ in a grand canonical ensemble of temperature $T$ and chemical potential $\mu$ via a classical Metropolis algorithm. 
For all simulations, we use $T= 40\, \mnK$  and adjust $\mu$ such that the total atom number $N$ is about $2 \times 10^4$.  
Each initial state is propagated using the classical equations of motion. 

To create a Josephson junction we add the term $\mathcal{H}_{ex} = \int \mathrm{d}{\bf r}\, V(x,t) n({\bf r}, t)$, 
where $n({\bf r}, t) = |\psi({\bf r} , t)|^2$ is the local density and $V(x, t)$ is the barrier potential of the form 
\begin{equation}\label{eq:pot} 
V(x,t)  = V_0 (t) \exp \Bigl[- \frac{ 2\bigl( x-x_0- x(t) \bigr)^2}{w^2} \Bigr]. 
\end{equation}
Here, $V_0(t)$ is the height and $w$ is the width of the barrier.  
The initial location of the barrier potential is fixed at $x_0= 32\, \mum$, while  $x(t)$ is the driving term. 
As the barrier size is at the limit of the experimental resolution, the experimental estimation of its height is particularly challenging. 
For this reason, we perform the numerical simulation by fixing the value of the barrier size to be compatible with the confidence range of the experimental one and scan $V_0/\mu_0$ to obtain the best agreement with the experimental data of AC and DC drive reported in Fig.~\ref{fig3}.
We choose $w=0.7\, \mum$, and ramp up $V_0(t)$ linearly to $V_0/\mu_0=1.45$ over $100\, \mms$ and then wait for $50\, \mms$ to achieve equilibrium, 
where $\mu_0$ is the chemical potential averaged over the $z$ direction. 
This creates a weak link at $x_0$ by separating the cloud into two subsystems 
(referred to as the left and right reservoirs).  
The driving term is given by  
\begin{align}
x(t) = v_{DC} t + x_{AC} \sin(\omega t),
\end{align}
where $v_{DC}$ is the DC barrier velocity, $x_{AC}$ is the AC driving amplitude and $\omega$ is the AC driving frequency \cite{Singh2023}. 
Similarly to the experiment, we write the current as 
$I(t) = I_{DC} + I_{AC} \cos(\omega t) $, where $ I_{DC} = v_{DC} I_c/v_c$ and $ I_{AC} = x_{AC} \omega I_c/v_c$, with $I_c$ being the critical current and $v_c$ the critical velocity.  
For various driving parameters, we analyze $I-\Delta \mu$ characteristic curves as a function of the bias current $I_{DC}$ after three driving periods. 
$\Delta \mu= \mu_R - \mu_L$ is the chemical-potential difference between the left ($\mu_L$) and right ($ \mu_R$) reservoir, which is determined using the total atom number in each reservoir. 
In Fig.~\ref{fig3}A of the main text, we show the simulation results of $I-\Delta \mu$ curves for undriven ($I_{AC}/I_c=0$) and driven ($I_{AC}/I_c=1.4$ and $\omega=2 \pi \times 175\, \mHz$) junctions.
We perform simulations of the AC response of the junction as a function of the driving frequency $\omega$ and report these results in Fig.~\ref{figSM_BEC_Shap}B, which are in excellent agreement with the experimental findings.
To determine the height of Shapiro steps we fit $I-\Delta \mu$ curves with sigmoid functions as in the experiments. 

\begin{figure}[t!]
\centering
\includegraphics[width=0.45
\textwidth]{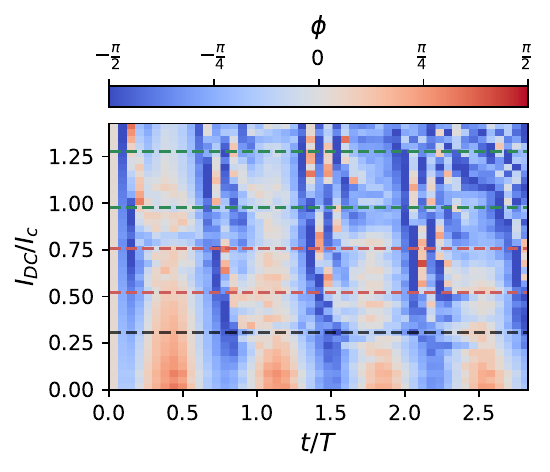}
\caption{\textbf{Phase dynamics of an AC-driven Josephson junction.}
Time evolution of the phase difference $\phi$ across the barrier for varying $I_{DC}/I_c$, as obtained from the numerical simulation under the same condition as for Fig.~\ref{fig3} of the main text ($\omega=2 \pi \times 175\, \mHz$, $I_{AC}/I_c=1.4$).
Dashed lines mark the edges of the 0-th (black), 1-st (red), and 2-nd (green) Shapiro step in the corresponding $I- \Delta \mu$ curve. The step edges are obtained as the current points where the sigmoid fit changes by 10\% from the plateau value.
}
\label{figS1}
\end{figure}

To determine the $\phi-I$ characteristic of the junction we calculate the phase difference in the vicinity of the barrier 
\begin{align}
\phi = \phi(x-2l) - \phi(x+2l),     
\end{align}
where $x(t)$ is the dynamic location of the barrier at time $t$. 
In Fig.~\ref{figS1} we show the time evolution $\phi(t)$, averaged over many samples, at varying $I_{DC}/I_c$ for the AC-driven system. From the phase change at the end of driving we obtain the $\phi-I$ curves, shown in Fig.~\ref{fig3}B of the main text.

\begin{figure*}[ht!]
\centering
\includegraphics[width=0.9\textwidth]{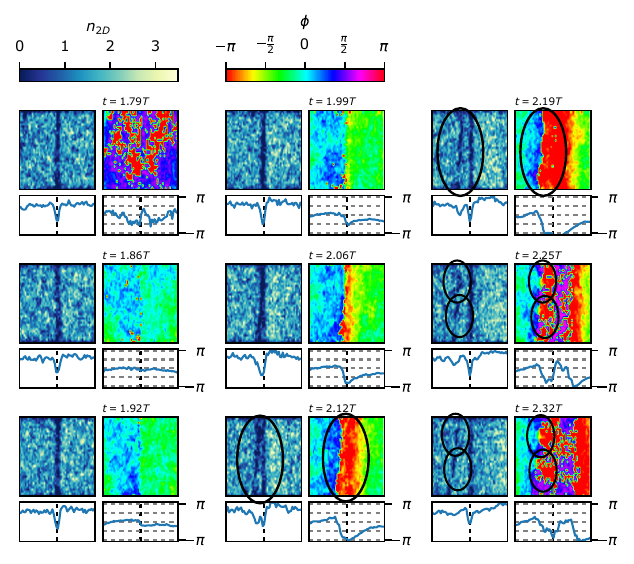}
\caption{\textbf{Phase slip process in linear junctions: second cycle}
Panels correspond to different times of evolution. Each panel is composed of: the density (top left), mean density profile (bottom left), phase (top right), and mean phase profile (bottom right).
}
\label{supFigVJsimMovie}
\end{figure*}

\begin{figure*}[ht!]
\centering
\includegraphics[width=0.9\textwidth]{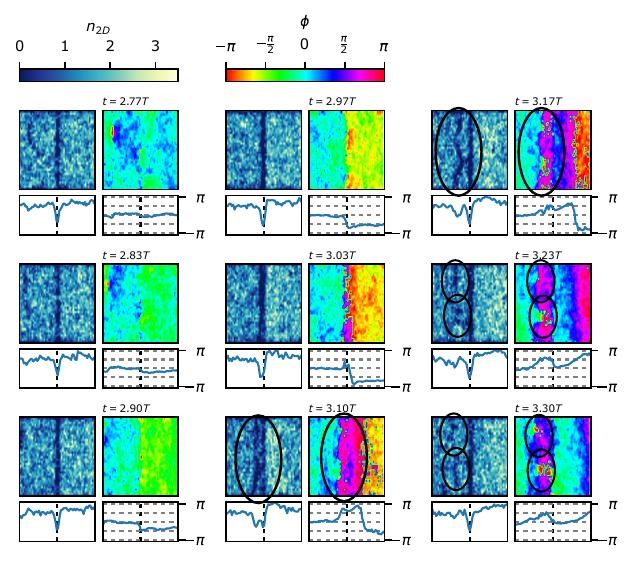}
\caption{\textbf{Phase slip process in linear junctions: third cycle}
Panels correspond to different times of evolution. Each panel is composed of: the density (top left), mean density profile (bottom left), phase (top right), and mean phase profile (bottom right).
}
\label{supFigVJsimMovie_c2}
\end{figure*}

\subsubsection*{Vortex dipole creation: phase slip process}

As mentioned in the main text, depending on the junction dimensionality, phase slips can manifest as different topological excitations \cite{Piazza2010, Tinkham2004, Binanti2021, Polo2018, Wlazowski2023, Xhani2020, Burchianti2018, Singh2023}, in particular for our experimental conditions as vortex-antivortex pair in two dimensions, and arise from a multi-step process as soliton decays into a different topological excitation \cite{Cetoli2013, Lombardi2017, VanAlphen2019, Ku2016}.
Our classical-field simulations corroborate this mechanism as shown in Fig.~\ref{supFigVJsimMovie}-\ref{supFigVJsimMovie_c2}, where a phase difference is accumulated along the extended junction at the end of the second (Fig.~\ref{supFigVJsimMovie}) and third cycle (Fig.~\ref{supFigVJsimMovie_c2}). After the solitonic excitation is released from the barrier, the soliton decays into multiple vortex-antivortex pairs possibly via a snake-instability \cite{Cetoli2013, Lombardi2017}. It is important to note that the depinning of the soliton and the subsequent decay process happens on a timescale faster than the modulation period. The critical wavevector at which the snake instability occurs near unitarity \cite{Lombardi2017}, and at unitarity reaches $k_c\sim 0.93 \,k_F$ and decreases in the BEC side of the crossover according to the formula $k_c \sim \sqrt{2}\xi_L^{-1}$. 
The critical wavelength is therefore $\lambda_c=2\pi/k_c = \sqrt{2}\pi \xi_L\approx 2.5(2) \mu $m in the BEC, and $\lambda_c=2\pi/k_c= 1.8(2)\ \mu$m in the UFG.
In particular, our junction transverse size is large compared to the critical length $L_y/ \lambda_c \approx 7$, allowing multiple "snake" oscillations along the junction transverse direction.

\subsection*{Shapiro steps in the BEC regime}

\begin{figure}[ht!]
\centering
\includegraphics[width=0.5
\textwidth]{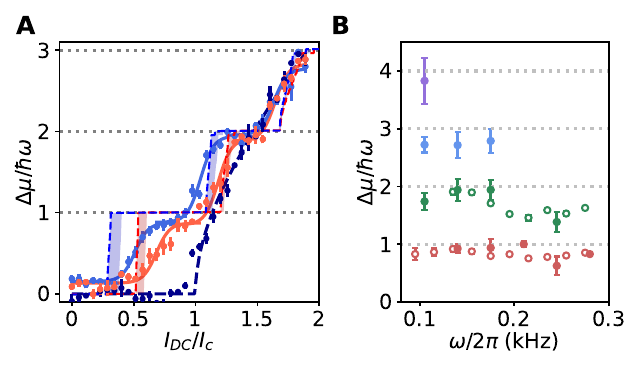}
\caption{\textbf{Shapiro Steps in a BEC}
(\textbf{A}) Current-voltage characteristic for a modulation frequency of $\omega=2\pi \times 175 \, $Hz, considering three different amplitudes of modulation: $x_{AC}=0$ (dark blue), $x_{AC}= 1.2 \, \mu$m (orange) and $x_{AC}= 1.6 \, \mu$m (royal blue). The dashed lines correspond to a fit with the stationary solution of the undriven overdamped RCSJ model. The continuous shaded curve shows the results of the RCSJ model for the driven scenario, where the shades account for the fitting error on $I_c$. The lines that join the experimental points represent the phenomenological fit of each step using a sigmoid function. (\textbf{B}) Shapiro step height characterization from measurements (filled symbols) and numerical simulations (open symbols). Each color represents different Shapiro steps: 1st (red), 2nd (green), 3rd (blue), and 4th (purple).}
\label{figSM_BEC_Shap}
\end{figure}

In this section, we report the characterization of the Shapiro step properties for the BEC junction ($1/k_Fa = \SI{3.3\pm0.1}{}$), together with a detailed description of the phenomenological fitting procedure, employed to analyze all the $I - \Delta \mu$ curves, independently from the interaction regime.
The DC current-voltage data are fitted with the analytical solution of the overdamped RCSJ model, namely with $\Delta \mu = \sqrt{ I_{DC}^2 - I_c^2}/G$ (dark blue dashed line of Fig.~\ref{figSM_BEC_Shap} A), keeping both $I_c$ and $G$ as free parameter. 
On the other hand, to quantitatively characterize the Shapiro step height and width, we fit the AC $I - \Delta \mu$ curve with as many independent sigmoid functions as the number of visible steps (solid lines in Fig.~\ref{figSM_BEC_Shap} A). In particular, we fit the $n$-th step with the function:
\begin{equation}
    \Delta \mu = \frac{\Delta \mu_{n}^{\mathrm{rel}}}{\left(1+ \exp{\left( -\frac{I_{DC}-I_n}{\Gamma}\right)})\right)}+ \Delta \mu_{n-1}^{\mathrm{rel}},
\end{equation}
where $\Delta \mu_{n}^{\mathrm{rel}}$ is the relative height of the $n$-th step and $I_n$ its position. From these parameters we compute the $n$-th step height as $\Delta \mu_{n} = \sum_{i=0}^{n} \Delta \mu_{i}^{\mathrm{rel}}$ and its width as $\Delta I_n = I_{n+1}-I_n$. The step height characterization as a function of the modulation frequency in the BEC regime is reported in Fig.~\ref{figSM_BEC_Shap}. Here, the experimental data (filled symbols) are compared with the results of classical-field numerical simulations (empty symbols), performed as described in the dedicated section, and analyzed with the same fitting procedure as described above. The numerical simulations are in very good agreement with the experimental results, both confirming the presence of Shapiro steps at $\Delta \mu = n \hbar \omega$ in the BEC regime in the explored range of frequency. In particular, in both cases, we observe a reduction of the measured step height for increasing modulation frequency, which could be due to the coupling of the drive with trap excitation along the $z$-direction. In fact, for AC drive frequencies approaching the vertical trap frequency, we expect the Shapiro dynamics to mix with trap excitations, possibly redistributing the injected energy between the two contributions. We observe a similar reduction also for the junction at unitarity at $\omega \simeq 2 \pi \times 500\, $Hz, compatible with this scenario given the higher trap frequency in this regime. As already mentioned above, a reduction of the step height at higher modulation frequency could also arise from the finite speed of propagation of the generated $\Delta \mu$ at the barrier, given by the speed of sound. For high $\omega$, the space propagated by the $\Delta \mu$ pulses during three modulation cycles can be shorter than the junction size, so that the measured $\Delta \mu$ can be lower than the stationary value. 

In Fig.~\ref{figSM_BEC_Shap} A, we report also the overdamped RCSJ numerical solutions for the AC $I - \Delta \mu$ (red and blue dashed lines), obtained by using the value of $I_c$ and $G$ as extracted from the fit of the DC curve. The numerical solutions qualitatively represent the measured steps, despite a smoother trend visible both in the experimental data and in numerical simulations, visible also in the data reported in Fig.~\ref{fig3} A of the main text. Such a discrepancy is most likely given by larger conductance at BEC ($h C = 6.5(6) \,$s), with respect to unitarity ($h C = 1.8(2) \,$s).
The conductance is here computed as the inverse of the charging energy $E_c = 1/C = \frac{\partial \mu_L}{N_L} + \frac{\partial \mu_R}{N_R} \simeq 4 \frac{\partial \mu_0}{N}$, and is therefore proportional to the compressibility of the gas. 
In the BEC regime, the larger compressibility, with respect to unitarity, allows for density excitations at higher contrast to propagate in the junction as a consequence of the AC drive, making the amplitude of the order parameters in the two reservoirs non-spatially homogeneous. As a result, the dynamic of the junction is no longer well describable in terms of uniquely its relative phase $\phi$, and the RCSJ model becomes less accurate. We note that density excitations are observable also in the unitary junction as a consequence of the AC drive (see Fig.~\ref{figSM_density_evolution}), but the strong interactions reduce their contrast and thus keep the overdamped RCSJ model a good approximation.

\end{document}